\documentstyle[12pt]{article}
\advance\textheight by \topskip
\begin{document}
\vspace{-0.2cm}
\begin{flushright}
UG-8/94\\
SU-ITP-94-19\\
QMW-PH-94-13\\
hep-th/9410230\\
\end{flushright}
\vspace{.5cm}

\begin{center}
{\large\bf DUALITY VERSUS SUPERSYMMETRY\\
\vskip .6 cm
AND COMPACTIFICATION }\\
\vskip 0.5 cm

{\bf Eric Bergshoeff ${}^a$\footnote{E-mail address:
{\tt bergshoe@th.rug.nl}},
Renata Kallosh ${}^b$\footnote{E-mail address:
{\tt kallosh@physics.stanford.edu}}
and Tom\'as Ort\'{\i}n${}^{c}$\footnote{E-mail address:
{\tt ortin@qmchep.cern.ch}}}
\vskip 0.05cm
${}^a$ Institute for Theoretical Physics, University of Groningen,\\
Nijenborgh 4, 9747 AG Groningen, The Netherlands\\
\vskip 0.4truecm
${}^b$Department of Physics, Stanford University,\\
Stanford   CA 94305, USA\\
\vskip .4truecm
${}^{c}$ Department of Physics, Queen Mary and Westfield College, \\
Mile End Road, London E1 4NS, U.K.
\end{center}


\vskip 0.6 cm
\centerline{\bf ABSTRACT}
\begin{quotation}

We study the interplay between T-duality, compactification and
supersymmetry.  We prove that when the original configuration has
unbroken space-time supersymmetries, the dual configuration also does if
a special condition is met: the Killing spinors of the original
configuration have to be independent on the coordinate which corresponds
to the isometry direction of the bosonic fields used for duality.
Examples of ``losers" (T-duals are not supersymmetric) and ``winners"
(T-duals are supersymmetric) are given.

\end{quotation}

\newpage


\section*{Introduction}

Target-space duality (T-duality) is a powerful tool for generating new
classical solutions of string theory.  It can be used in the sigma model
context to generate new exact solutions but also in the context of the
leading order in $\alpha^{\prime}$ effective action to generate new
solutions to the low energy equations of motion.  Some of these
solutions have unbroken supersymmetries.  The purpose of this paper is
to study the generic relation between the supersymmetric properties of
the original configuration and the dual one in the context of the low
energy effective action.

It has been observed that in some cases T-duality preserves unbroken
supersymmetry.  Well-known example are the supersymmetric string wave
solutions (SSW) \cite{kn:BKO} and their partners, dual waves
\cite{kn:BEK}, which in particular include fundamental string solutions
\cite{kn:DGHR}.  Another example of T-dual partners with unbroken
supersymmetries is given by a special class of fivebrane solutions
\cite{kn:CHS}, multimonopoles \cite{kn:Kh} and their duals, a special
class of stringy ALE instantons \cite{kn:Bian} which has a multicenter
metric.

The preservation of unbroken supersymmetries by duality is related in
principle to the fact that T-duality is just one of the {\it hidden
symmetries} of the supergravity theory that arises after dimensional
reduction \cite{kn:Schdual}.  These hidden symmetries are indeed
consistent with the supersymmetry of the dimensionally reduced theory.
However, some recently discovered counterexamples seem to contradict
this preliminary understanding.  Therefore, one of the main goals of our
analysis is to find the general condition that guarantees the
preservation of unbroken supersymmetries that it is not satisfied by
these counterexamples.  We will perform this analysis in the context of
$N=1,d=10$ supergravity without vector fields.  More general results
involving Abelian and non-Abelian vector fields and higher order
$\alpha^{\prime}$ corrections will be reported elsewhere \cite{kn:BKO2}.
Some of the results presented in this paper were announced in
\cite{kn:O2}.

The first counterexample known to us appears in a very simple case.  We
have found some time ago\footnote{It was suggested to us by A. Tseytlin
to check whether supersymmetry is preserved by duality in this case.}
that if one starts with ten-dimensional flat space (which has all
supersymmetries unbroken) in polar coordinates and performs a
$T$-duality transformation with respect to the angular coordinate
$\varphi$, the resulting configuration has no unbroken supersymmetries
whatsoever.

The explanation of this apparently inconsistent situation will be found
in a Kaluza-Klein-type analysis of the fermionic supersymmetry
transformation rules of $N=1,d=10$ supergravity.  In the conventional
dimensional reduction of this theory by compactification of one
dimension (with coordinate $x$, say) one only considers those field
configurations that do not depend on the compact coordinate, and one
only considers those supersymmetry transformations generated by
parameters $\epsilon$ that are independent of $x$ as well, projecting
the rest out of the resulting $N=1,d=9$ theory which is the case $n=1$
of Ref.~\cite{kn:GNS}.  If the Killing spinor of the ten-dimensional
configuration depends on $x$, the configuration will not be
supersymmetric in nine dimensions.  The effect of the nine-dimensional
hidden symmetries in the ten-dimensional supersymmetry is unknown, while
in the nine-dimensional theory is just a $O(1,1)$ group completely
consistent with supersymmetry \cite{kn:GNS}.  This is exactly what
happens in the counterexample above: the Killing spinor depends on
$\varphi$ when a $\varphi$-independent frame is used.

Recently Bakas \cite{kn:Bakas} has found a more interesting example of
the loss of unbroken supersymmetries after a series of T- and S-duality
transformations, T-duality being the responsible of this loss.  In his
scheme supersymmetry is lost if the Killing vector with respect to which
one dualizes has not self-dual covariant derivatives.  We believe that
his example also satisfies our criterion: if one calculated explicitly
the Killing spinors of such configurations\footnote{To apply our
criterion one has to find the Killing spinors explicitly.} in an
$x$-independent frame, they would depend on the isometry direction $x$.
We hope these different criteria can be shown to be equivalent for
these configurations.

This article is structured as follows: In Section~\ref{sec-DD-1} we set
up the general problem of dimensionally reducing one dimension in the
low-energy string effective action in absence of gauge fields, mainly
for fixing the conventions and notation.  We describe the effect of
T-duality on the compactified dimension from the point of view of the
lower dimensional theory.  In Section~\ref{sec-susy} we study the effect
of T-duality on the supersymmetry properties of purely bosonic
configurations of the zero-slope limit of heterotic string theory in ten
dimensions.  Accordingly we investigate the behavior under T-duality of
the supersymmetry rules of pure $N=1$, $d=10$ supergravity.  Using a
Kaluza-Klein basis of zehnbeins we rewrite the ten-dimensional
supersymmetry transformation rules in a manifestly T-duality-invariant
form for configurations which have unbroken supersymmetries with the
Killing spinor independent on the isometry direction.  In
Section~\ref{sec-examples} we present examples of configurations with
(broken) unbroken supersymmetry after duality in accordance with
(dependence) independence of the Killing spinor on isometry direction.
Section~\ref{sec-conclusion} contains our conclusions.  Finally,
Appendix~\ref{sec-truncation} contains some additional results of our
work:  we dimensionally reduce $N=1$ $d=10$ supergravity to $d=4$ and
study the truncation of the lower dimensional theory consisting in
setting to zero all the fields which are {\it matter} from the point of
view of $N=4,d=4$ supergravity.  The remaining fields are found to be
duality-invariant.  Therefore, when a supersymmetric compactification is
done and the resulting theory is truncated to pure supergravity,
T-duality in the compactified directions has no effect whatsoever on the
theory.


\section{From $D$ to $D-1$ dimensions}
\label{sec-DD-1}

The D-dimensional action we start from is

\begin{equation}
S=\frac{1}{2}\int d^{D}x
\sqrt{-\hat{g}} e^{-2\hat{\phi}} [-\hat{R}
+4(\partial\hat{\phi})^{2} -\frac{3}{4}\hat{H}^{2}]\, ,
\label{eq:actionD1}
\end{equation}
where the fields are the metric, the axion and the dilaton
$\{\hat{g}_{\hat{\mu}\hat{\nu}},\hat{B}_{\hat{\mu}\hat{\nu}},
\hat{\phi}\}$ and our conventions are those of Ref.~\cite{kn:BEK}.  In
particular the axion field strength $\hat{H}$ is given by
\begin{equation}
\hat{H}_{\hat{\mu}\hat{\nu}\hat{\rho}} =
\partial_{[\hat{\mu}}\hat{B}_{\hat{\nu}\hat{\rho}]}\, .
\label{eq:H1}
\end{equation}
All the D-dimensional entities carry a hat and the D-1 dimensional ones
don't. Then the indices take the values
\begin{equation}
\hat{\mu}=(0,\ldots,D-2,D-1)=(\mu,D-1)\, .
\end{equation}

We call the coordinate $x^{D-1}\equiv x$.  To distinguish between curved
and flat indices when confusion may arise, we underline the curved ones
($\xi^{\underline{x}}$, for instance).  Now we assume that the fields
are independent of the coordinate $x$, i.e. there exists a Killing
vector $\hat{k}^{\hat{\mu}}$ such that
\begin{equation}
\hat{k}^{\hat{\mu}}\partial_{\hat{\mu}}=\partial_{\underline{x}}\, .
\end{equation}
Then, in this coordinate system, the components of the Killing vector
are
\begin{equation}
\hat{k}^{\hat{\mu}}=\delta^{\hat{\mu} \underline{x}}\,
,\hspace{.5cm}\hat{k}_{\hat{\mu}}=\hat{g}_{\hat{\mu} \underline{x}}\, ,
\hspace{.5cm}\hat{k}^{2}=\hat{k}^{\hat{\mu}}\hat{k}_{\hat{\mu}}=
\hat{g}_{\underline{x}\underline{x}}\, ,
\end{equation}
and the metric can be rewritten as follows
\begin{equation}
ds^{2}=\hat{k}^{-2}(\hat{k}_{\hat{\mu}}dx^{\hat{\mu}})^{2}
+(\hat{g}_{\mu\nu}-
\hat{k}^{-2}\hat{k}_{\mu}\hat{k}_{\nu})dx^{\mu}dx^{\nu}\, .
\end{equation}

The Killing vector can be either time-like or space-like, but not null.
We will keep our expressions valid for both cases because from the point
of view of T-duality both are equally interesting \cite{kn:W} and the
compactification of a time-like coordinate is not usually considered in
the literature because it gives rise to an inconsistent lower
dimensional theory.  We consider here the lower dimensional theory just
as a tool.

The above action enjoys invariance under the following Buscher's
\cite{kn:B} T-duality transformations
\begin{equation}
\begin{array}{rclrcl}
\tilde{\hat{g}}_{\underline{x}\underline{x}} & = &
1/\hat{g}_{\underline{x}\underline{x}}\, , &
\tilde{\hat{B}}_{\underline{x}\mu} & = &
\hat{g}_{\underline{x}\mu}/\hat{g}_{\underline{x}\underline{x}}\, ,
\\
& & & & &
\\
\tilde{\hat{g}}_{\underline{x}\mu} & = &
\hat{B}_{\underline{x}\mu}/\hat{g}_{\underline{x}\underline{x}}\, , &
\tilde{\hat{B}}_{\mu\nu} & = &
\hat{B}_{\mu\nu}+(\hat{g}_{\underline{x}\mu}
\hat{B}_{\nu \underline{x}}-
\hat{g}_{\underline{x}\nu}\hat{B}_{\mu \underline{x}})
/\hat{g}_{\underline{x}\underline{x}}\, ,
\\
& & & & &
\\
\tilde{\hat{g}}_{\mu\nu} & = &
\hat{g}_{\mu\nu}-(\hat{g}_{\underline{x}\mu}\hat{g}_{\underline{x}\nu}-
\hat{B}_{\underline{x}\mu}\hat{B}_{\underline{x}\nu})
/\hat{g}_{\underline{x}\underline{x}}\, ,&
\tilde{\hat{\phi}} & = & \hat{\phi} -\frac{1}{2}\log
|\hat{g}_{\underline{x}\underline{x}}|\, .
\end{array}
\label{eq:Buscher}
\end{equation}
Checking directly the invariance of the action Eq.~(\ref{eq:actionD1})
under the above transformations is a very involved calculation but if we
compactify the redundant coordinate $x$, checking duality will be
very easy.

Now we are going to dimensionally reduce the above action to $D-1$
dimensions by compactifying the redundant dimension $x$. We use the
standard techniques of Scherk and Schwarz \cite{kn:SSC}. First we
parametrize the $D$-bein as follows
\begin{equation}
(\hat{e}_{\hat{\mu}}{}^{\hat{a}})=
\left(
\begin{array}{cc}
e_{\mu}{}^{a} & k A_{\mu} \\
0           & k         \\
\end{array}
\right)
\, ,
\hspace{1cm}
(\hat{e}_{\hat{a}}{}^{\hat{\mu}})=
\left(
\begin{array}{cc}
e_{a}{}^{\mu} & -A_{a} \\
0           & k^{-1}   \\
\end{array}
\right)\, ,
\label{eq:basis}
\end{equation}
where
\begin{equation}
k=|\hat{k}_{\hat{\mu}}\hat{k}^{\hat{\mu}}|^{\frac{1}{2}}\, ,
\end{equation}
and $A_{a}=e_{a}{}^{\mu}A_{\mu}$.  The functions
$\hat{e}_{\hat{\mu}}{}^{\hat{a}}$ do not depend on $x$.  Note that
$\hat{k}_{\hat{\mu}}\hat{k}^{\hat{\mu}}=\hat{\eta}_{xx}k^{2}$.  With our
conventions (mostly minuses signature) $\hat{\eta}_{xx}$ is positive if
$x$ is a time-like coordinate and $\hat{k}^{\hat{\mu}}$ a time-like
vector, and $\hat{\eta}_{xx}$ is negative if $x$ and
$\hat{k}^{\hat{\mu}}$ are both space-like.

With this parametrization, the D-dimensional fields decompose as follows
\begin{equation}
\begin{array}{rclrcl}
\hat{g}_{\underline{x}\underline{x}} & = & \hat{\eta}_{xx}k^{2}\, ,&
\hat{B}_{\underline{x}\mu} & = & B_{\mu}\, ,
\\
& & & & &
\\
\hat{g}_{\underline{x}\mu} & = & \hat{\eta}_{xx}k^{2}A_{\mu}\, ,&
\hat{B}_{\mu\nu} & = & B_{\mu\nu}+A_{[\mu}B_{\nu]}\, ,
\\
& & & & &
\\
\hat{g}_{\mu\nu} & = & g_{\mu\nu}+\hat{\eta}_{xx}k^{2}A_{\mu}A_{\nu}\,
,\hspace{1.5cm}& \hat{\phi} & = & \phi +\frac{1}{2}\log k\, ,
\end{array}
\end{equation}
where $\{ g_{\mu\nu},B_{\mu\nu},A_{\mu},B_{\mu},k,\phi\}$ are the
$(D-1)$-dimensional fields. They are given in terms of the
$D$-dimensional fields by
\begin{equation}
\begin{array}{rclrcl}
g_{\mu\nu} & = & \hat{g}_{\mu\nu}-\hat{g}_{\underline{x}\mu}
\hat{g}_{\underline{x}\nu}/\hat{g}_{\underline{x}\underline{x}}\, ,&
B_{\mu} & = & \hat{B}_{\underline{x}\mu}\, ,
\\
& & & & &
\\
B_{\mu\nu} & = & \hat{B}_{\mu\nu}+
\hat{g}_{\underline{x}[\mu}\hat{B}_{\nu]\underline{x}}
/\hat{g}_{\underline{x}\underline{x}}\, ,\hspace{1.5cm}&
\phi & = & \hat{\phi}-\frac{1}{4}
\log|\hat{g}_{\underline{x}\underline{x}}|\, ,
\\
& & & & &
\\
A_{\mu} & = & \hat{g}_{\underline{x}\mu}
/\hat{g}_{\underline{x}\underline{x}}\, ,&
k & = & |\hat{g}_{\underline{x}\underline{x}}|^{\frac{1}{2}}\, .
\end{array}
\label{eq:fields}
\end{equation}
Then the $D$-dimensional action Eq.~(\ref{eq:actionD1}) is identically
equal to
\begin{eqnarray}
S & = &\frac{1}{2}\int d^{(D-1)}x  \sqrt{\eta_{xx}g} e^{-2\phi}
[-R+4(\partial\phi)^{2}-\frac{3}{4}H^{2}-
\nonumber \\
& &
\nonumber \\
& - &(\partial\log k)^{2} -\hat{\eta}_{xx}\frac{1}{4}k^{2}F^{2}(A)
-\hat{\eta}_{xx}\frac{1}{4}k^{-2}F^{2}(B)]\, ,
\label{eq:actionD-11}
\end{eqnarray}
where
\begin{eqnarray}
F_{\mu\nu}(A) & = & 2\partial_{[\mu}A_{\nu]}\, ,
\hspace{1cm}
F_{\mu\nu}(B) = 2\partial_{[\mu}B_{\nu]}\, ,
\nonumber \\
& &
\nonumber \\
H_{\mu\nu\rho} & = & \partial_{[\mu}B_{\nu\rho]}+
\frac{1}{2}A_{[\mu}F_{\nu\rho]}(B)
+\frac{1}{2}B_{[\mu}F_{\nu\rho]}(A)\, ,
\end{eqnarray}
are the vectors and antisymmetric tensor field strengths.

Eq.~(\ref{eq:actionD-11}) can be interpreted as a $(D-1)$-dimensional
action for the above $(D-1)$-dimensional fields.  Observe that, when $x$
is a time-like coordinate, the vector fields kinetic terms have the
wrong signs in the above action.

Now, using first the definitions of the $(D-1)$-dimensional fields in
terms of the $D$-dimensional ones Eqs.~(\ref{eq:fields}) and Buscher's
duality rules Eqs.~(\ref{eq:Buscher}), it is very easy to check that the
duals of the $(D-1)$-dimensional fields are:
\begin{equation}
\begin{array}{rclrcl}
\tilde{g}_{\mu\nu} & = & g_{\mu\nu}\, ,&
\tilde{A}_{\mu} & = & B_{\mu}\, ,
\\
& & & & &
\\
\tilde{B}_{\mu\nu} & = & B_{\mu\nu}\, ,\hspace{2cm}&
\tilde{B}_{\mu} & = & A_{\mu}\, ,
\\
& & & & &
\\
\tilde{\phi} & = & \phi\, ,&
\tilde{k} & = & k^{-1}\, ,
\\
& & & & &
\\
\end{array}
\label{eq:dualreduced}
\end{equation}
that is, in the $(D-1)$-dimensional theory the only effect of
T-duality is to interchange the vector fields $A_{\mu}$ and
$B_{\mu}$ and to invert $k$. This is an obvious symmetry of the
$(D-1)$-dimensional action Eq.~(\ref{eq:actionD-11}) which, on the
other hand is identically equal to the $D$-dimensional one
Eq.~(\ref{eq:actionD1}). From the lower dimensional point of view,
the invariance of the action under T-duality is manifest.

Observe that, in particular, the $D-1$-beins are duality-invariant. This
is completely consistent with the transformation rules derived in
Ref.~\cite{kn:BEK}
\begin{eqnarray}
\tilde{\hat{e}}_{\underline{x}}{}^{\hat{a}} & = &
\frac{1}{\hat{g}_{\underline{x}\underline{x}}}
\hat{e}_{\underline{x}}{}^{\hat{a}}\, ,
\nonumber \\
& &
\nonumber \\
\tilde{\hat{e}}_{\mu}{}^{\hat{a}} & = &
\mp\hat{e}_{\mu}{}^{\hat{a}}
-\frac{1}{\hat{g}_{\underline{x}\underline{x}}}
(\hat{g}_{\underline{x}\mu} \pm\hat{B}_{\underline{x}\mu})
\hat{e}_{\underline{x}}{}^{\hat{a}}\, ,
\end{eqnarray}
and it is this property of the Kaluza-Klein basis Eq.(\ref{eq:basis})
which simplifies the transformation rules is justifies its use here.

Of course, the transformation one sees in the lower dimensional theory
is part of the $O(d,d)$ symmetry exhibited in
Ref.~\cite{kn:Schdual} when one compactifies $d$ dimensions.  Now
we have made this relation very explicit and it is going to be extremely
useful for the study of the unbroken supersymmetries of the dual
configurations.


\section{Duality versus Supersymmetry}
\label{sec-susy}

In this section we investigate the general relation between unbroken
supersymmetries before and after a T-duality transformation using the
results of the previous Section with $D=10$.  Specifically we are going
to analyze the effect of a T-duality transformation on $N=1,d=10$
supergravity Killing spinors.  To do this one needs to know how the
zehnbeins transform under duality.  As we explained in the previous
section, the zehnbein duality transformation laws were found in
Ref.~\cite{kn:BEK} and reduce to Eqs.~(\ref{eq:dualreduced}) for the
$x$-independent Kaluza-Klein basis of zehnbeins Eq.~(\ref{eq:basis})
where a clear distinction between the cases in which unbroken
supersymmetry is preserved and those in which it is not arises
naturally.

We consider here the zero slope limit of the heterotic string theory
without gauge fields, which is given by $N=1,d=10$ supergravity.  The
bosonic part of the action of $N=1$, $d=10$ supergravity in absence of
vector fields is given by Eq.~(\ref{eq:actionD1}) with $D=10$:

\begin{equation}
S=\frac{1}{2}\int d^{10}x
\sqrt{-\hat{g}} e^{-2\hat{\phi}} [-\hat{R}
+4(\partial\hat{\phi})^{2} -{\textstyle\frac{3}{4}}\hat{H}^{2}]\, ,
\label{eq:action10}
\end{equation}
with $H$ given by Eq.~(\ref{eq:H1}). The corresponding fermionic
supersymmetry transformation rules are
\begin{eqnarray}
\delta_{\hat{\epsilon}}\hat{\psi}_{\hat{c}} & = & [\partial_{\hat{c}}
-{\textstyle\frac{1}{4}}(\hat{\omega}_{\hat{c}}{}^{\hat{a}\hat{b}}
-{\textstyle\frac{3}{2}}\hat{H}_{\hat{c}}{}^{\hat{a}\hat{b}})
\hat{\gamma}_{\hat{a}\hat{b}}]\hat{\epsilon}\, ,
\nonumber \\
& &
\nonumber \\
\delta_{\hat{\epsilon}}\hat{\lambda} & = &
(\hat{\gamma}^{\hat{c}}\partial_{\hat{c}}\hat{\phi}
+{\textstyle\frac{1}{4}}\hat{H}_{\hat{a}\hat{b}\hat{c}}
\hat{\gamma}^{\hat{a}\hat{b}\hat{c}})\hat{\epsilon}\, .
\label{eq:susyrules10}
\end{eqnarray}

Now we assume that some spinor $\hat{\epsilon}$ makes these equations
vanish (i.e. $\hat{\epsilon}$ is a Killing spinor\footnote{Actually a
spinor that makes Eqs.~(\ref{eq:susyrules10}) vanish needs to have a
specific asymptotic behavior in order to be a Killing spinor, but these
details will not concern us in this discussion.}) for some specific
$x$-independent field configuration and we want to investigate whether
this $\hat{\epsilon}$ is also a Killing spinor of the T-dual field
configuration or whether it is related to another Killing spinor of the
dual field configuration, as in the S-duality case \cite{kn:O}.  To
investigate this problem we rewrite the above equations in terms of the
nine-dimensional fields.  It is perhaps worth stressing that we are not
reducing the ten-dimensional gamma matrices nor the ten-dimensional
spinors, which are the objects we are interested in.  Again, here,
dimensional reduction can be understood as a tool for having under
control the duality transformations.  All the indices used below are
flat.  We also use the slightly unusual notation (observe that the gamma
matrices are ten-dimensional but the indices contracted are the
nine-dimensional ones)
\begin{equation}
\not\!\partial\equiv\hat{\gamma}^{a}\partial_{a}\, ,
\hspace{.5cm}
\not\!F\equiv \hat{\gamma}^{ab}F_{ab}\, ,
\hspace{.5cm}
\not\!F_{a}\equiv \hat{\gamma}^{b}F_{ab}\, ,
\hspace{.5cm}
\not\!\!H\equiv \hat{\gamma}^{abc}H_{abc}\, .
\end{equation}

We get for the $x$-component (flat) and the $a$-component of the
gravitino transformation and for the dilatino transformation
\begin{eqnarray}
\hat{\gamma}_{x}\delta_{\hat{\epsilon}}\hat{\psi}_{x} & = &
\{k^{-1} \partial_{\underline{x}} -{\textstyle\frac{1}{8}}
[\hat{\eta}_{xx}k\not\!F(A) +k^{-1}\not\!F(B)]
+{\textstyle\frac{1}{2}} \hat{\gamma}_{x} (\not\!\partial\log k) \}
\hat{\epsilon}=0\, ,
\nonumber \\
& &
\nonumber \\
\delta_{\hat{\epsilon}}\hat{\psi}_{a} & = &
\{[\partial_{a} -{\textstyle\frac{1}{4}}(\omega_{a}{}^{bc}
-{\textstyle\frac{3}{2}}H_{a}{}^{bc})\hat{\gamma}_{bc}]
-{\textstyle\frac{1}{8}}\hat{\gamma}_{x} [\not\!F_{a}(A)
-\hat{\eta}_{xx} \not\!F_{a}(B)]
-A_{a}\partial_{\underline{x}}\}\hat{\epsilon}=0\, ,
\nonumber \\
& &
\nonumber \\
\delta_{\hat{\epsilon}}\hat{\lambda} & = &
\{\not\!\partial\phi
+{\textstyle\frac{1}{4}}\not\!\!H
-{\textstyle\frac{1}{4}}\hat{\eta}_{xx}k^{-1} \hat{\gamma}_{x}
\not\!F(B) +{\textstyle\frac{1}{2}}\not\!\partial\log
k\}\hat{\epsilon}=0\, ,
\end{eqnarray}
respectively.

As they stand, none of these equations is separately manifestly
duality-invariant.  Unless we assume in what follows that the Killing
spinor of the original configuration does not depend on the isometry
direction $x$, no further progress can be made in relating the
supersymmetry of the original configurations with that of the final one.
Thus we require that
\begin{equation}
\partial_{\underline{x}} \hat{\epsilon}=0\, .
\label{eq:condition}
\end{equation}

Using this assumption we have the Killing spinor equations in the form
\begin{eqnarray}
\{{\textstyle\frac{1}{8}}
[\hat{\eta}_{xx}k\not\!F(A) +k^{-1}\not\!F(B)]
-{\textstyle\frac{1}{2}} \hat{\gamma}_{x} (\not\!\partial\log k) \}
\hat{\epsilon} & = & 0\, ,
\nonumber \\
& &
\nonumber \\
\{[\partial_{a} -{\textstyle\frac{1}{4}}(\omega_{a}{}^{bc}
-{\textstyle\frac{3}{2}}H_{a}{}^{bc})\hat{\gamma}_{bc}]
-{\textstyle\frac{1}{8}}\hat{\gamma}_{x} [\not\!F_{a}(A)
-\hat{\eta}_{xx} \not\!F_{a}(B)]\}\hat{\epsilon} & = & 0\, ,
\nonumber \\
& &
\nonumber \\
\{\not\!\partial\phi +{\textstyle\frac{1}{4}}\not\!\!H
-{\textstyle\frac{1}{4}}\hat{\eta}_{xx}k^{-1} \hat{\gamma}_{x}
\not\!F(B)
+{\textstyle\frac{1}{2}}\not\!\partial\log k\}\hat{\epsilon} & = & 0\, ,
\end{eqnarray}

Still, after assuming $\partial_{\underline{x}}\hat{\epsilon}=0$, not
all of the Killing spinor equations are manifestly separately duality
invariant.  To be precise (and here we take $\hat{\eta}_{xx}=-1$) using
the nine-dimensional version of Buscher's duality rules
Eq.~(\ref{eq:dualreduced}) the first and the second are
duality-invariant but the third is clearly not.  However, since by
assumption all of them are satisfied by $\hat{\epsilon}$, we are allowed
to combine them.  If we substitute the first into the third, we get the
following duality-invariant equation:

\begin{equation}
\{\not\!\partial\phi +{\textstyle\frac{1}{4}}\not\!\!H
+{\textstyle\frac{1}{8}}\hat{\gamma}_{x} [k\not\!F(A)
+k^{-1}\not\!F(B)]\}\hat{\epsilon}=0\, .
\end{equation}
This proves that $\hat{\epsilon}$ is a Killing spinor of the dual
configuration if it is so for the original configuration, that is
\begin{equation}
\tilde{\hat{\epsilon}}=\hat{\epsilon}\, .
\end{equation}

If we take $\hat{\eta}_{xx}=+1$ (time-like duality) it is easy to see
that
\begin{equation}
\tilde{\hat{\epsilon}}=\hat{\gamma_{x}}\hat{\epsilon}\, .
\label{eq:dualk2}
\end{equation}

Examples of these results  will be  discussed in
Section~\ref{sec-examples}.

The set of T-duality invariant supersymmetry equations that we have
generated by dimensional reduction should be nothing but the explicitly
$O(1,1)$-invariant $N=1,d=9$ supergravity theory of Ref.~\cite{kn:GNS}
for the case $n=1$ and in stringy frame, although the dimensional
reduction of the supersymmetry parameters, gamma matrices etc. still has
to be done.  There are factors of $e^{\phi}$ relating the Einstein-frame
and string-frame spinors too and the comparison between our results ond
those or Ref.~\cite{kn:GNS} is not straightforward.  It is clear,
however, that the correspondence disappears if we don't impose
Eq.~(\ref{eq:condition}) to the supersymmetry parameters.

We would like to stress that we have derived the condition of
preservation of unbroken supersymmetry Eq.~(\ref{eq:condition}) using
heavily a zehnbein basis of the form Eq.~(\ref{eq:basis}).  However,
after deriving this condition in that special frame we may ask ourselves
to which extent this condition is frame-dependent.  The answer is that
the same criterion is valid in any $x$-independent frame.  Indeed, if
one changed from the $x$-independent Kaluza-Klein frame discussed above
to any other $x$-independent frame, the Lorentz rotation involved would
not change the fact that the spinor is or is not $x$-dependent since the
same $x$-independent parameter $\hat{\omega}^{\hat{a}\hat{b}}$ appears
in the spinors and frames transformation laws $\hat{\epsilon}^{\prime} =
\exp{(\frac{1}{4} \hat{\omega}^{\hat{a}\hat{b}}
\hat{\gamma}_{\hat{a}\hat{b}})} \hat{\epsilon}$ and $\hat{e}_{\hat{\mu}}
= \exp{(\frac{1}{4} \hat{\omega}^{\hat{a}\hat{b}}
\hat{M}_{\hat{a}\hat{b}})} \hat{e}_{\hat{\mu}}$ where the
$\hat{M}_{\hat{a}\hat{b}}$s are the generators of the
ten-dimensional Lorentz group in the vector representation.


\section{Examples}
\label{sec-examples}

In this Section we are going to study  examples of supersymmetric
configurations and duality transformations which illustrate the
results of Section~\ref{sec-susy}.

1. {\bf Losers}: configurations that lose their unbroken supersymmetries
after T-duality:

\newcounter{eso1}
\begin{list}%
{\roman{eso1})}{\usecounter{eso1}
\setlength{\rightmargin}{\leftmargin}}

\item Our first example is flat ten-dimensional space-time in
      polar coordinates $\rho^{2} =(x^{1})^{2} +(x^{2})^{2},\,\,
      \tan{\varphi} =x^{2}/x^{1}$:
\begin{equation}
ds^{2}=dt^{2}-d\rho^{2}-\rho^{2}d\varphi^{2}-dx^{I}dx^{I}\, ,
\hspace{1cm}
I=3,\ldots,9\, .
\end{equation}
This solution of $N=1,d=10$ supergravity has all supersymmetries
unbroken. In the zehnbein basis
\begin{equation}
e_{t}{}^{0}=1,\, e_{\rho}{}^{1}=1,\, e_{\varphi}{}^{2}=\rho,\,
e_{\underline{I}}{}^{J}=\delta_{\underline{I}}{}^{J}\, ,
\end{equation}
which is of the type of that in Eq.~(\ref{eq:basis}), the Killing
spinors are given by
\begin{equation}
\epsilon=e^{-\frac{1}{4}\gamma_{1}\gamma_{2}\varphi}\epsilon_{0}\, ,
\end{equation}
where $\epsilon_{0}$ is a completely arbitrary constant
spinor\footnote{In Cartesian coordinates and in the most obvious frame
$\hat{e}_{\hat{\mu}}{}^{\hat{a}} =\delta_{\hat{\mu}}{}^{\hat{a}}$ the
Killing spinors are just arbitrary constant spinors and so have the
right asymptotic behavior.}.

After a duality transformation in the direction $\varphi$, we get the
solution
\begin{eqnarray}
ds^{2} & = & dt^{2}-d\rho^{2}-\rho^{-2}d\varphi^{2}-dx^{I}dx^{I}\, ,
\nonumber \\
& &
\nonumber \\
\phi & = & -\log \rho\, .
\end{eqnarray}
The dilatino supersymmetry rule implies that the Killing spinors of this
solution have to satisfy the constraint
\begin{equation}
\gamma^{1}\epsilon=0\, ,
\end{equation}
which can only be satisfied by $\epsilon=0$. Therefore, all the
supersymmetries of the dual solution of Minkowski space are broken. As
we saw in Section~\ref{sec-susy} this is related to the dependence of
the Killing spinors on $\varphi$ when we use adapted coordinates and
a $\varphi$-independent frame.

\item Our second example is the one recently found by Bakas in
Ref.~\cite{kn:Bakas}. He studied self-dual Euclidean metrics admitting a
Killing vector associated to the coordinate $\tau$, which generally can
be written in the form
\begin{equation}
ds^2 = V(d\tau + \omega_i dx^i)^2 + V^{-1} \gamma_{ij} dx^i dx^j\, .
\label{eq:bakas}
\end{equation}
Self-duality of the metric is an integrability condition for the
existence of unbroken supersymmetries.  What was actually observed in
\cite{kn:Bakas} was the breaking of the self-duality condition of the
configuration after the T-S-T chain of duality transformations.

The violation of supersymmetry in this example could be attributed to
T-duality, since, as we have said, S-duality is perfectly consistent
with supersymmetry.  Furthermore, the violation of supersymmetry by
T-duality was related to the nature of the Killing vector: T-duality
with respect to ``translational" Killing vectors would not violate
supersymmetry while T-duality with respect to ``rotational" Killing
vectors would.  In particular, this criterion was sufficient to show
that for configurations with flat three-dimensional metrics
$\gamma_{ij}= \delta_{ij}$ no violation of supersymmetry happened.
However, for some special choices of non-flat $\gamma_{ij}$ the
self-duality of the final configuration was violated.

{}From our point of view, this gives an interesting example of our general
statement that unless the Killing spinor in Kaluza-Klein basis is shown
to be independent on duality direction there is no reason to expect the
preservation of supersymmetry by T-duality.  A preliminary study shows
that all the cases found in Ref.~\cite{kn:Bakas} to violate
supersymmetry suffer from the problem of dependence of the Killing
spinor on the coordinate associated to the isometry. Observe that one of
his examples with $V=1$ and $\gamma_{ij}\neq\delta_{ij}$ is provided by
the case i) above.

\end{list}

2. {\bf  Winners}: configurations with unbroken supersymmetries that are
preserved by T-duality. Alternatively we could refer to them as those
configurations with unbroken supersymmetries and $x$-independent Killing
spinors since the results of Section~\ref{sec-susy} guarantee, without
the need of further proof, the supersymmetry of the dual configurations.

\newcounter{eso2}
\begin{list}%
{\roman{eso2})}{\usecounter{eso2}
\setlength{\rightmargin}{\leftmargin}}

\item The first example is provided by the SSW solutions and
the GFS solutions which are both supersymmetric and are known to be
related by duality in the direction $x$ \cite{kn:BKO,kn:BEK,kn:DGHR}.
Let us describe briefly these two classes of solutions. The SSW
solutions are
\begin{eqnarray}
\hat{ds}^{2} & = &
2dudv+2{\cal A}_{u}du^{2}
+2{\cal A}_{\underline{i}}dx^{\underline{i}}du
-dx^{\underline{i}}dx^{\underline{i}}\, ,
\nonumber \\
& &
\nonumber \\
\hat{B} & = & 2{\cal A}_{\underline{i}}dx^{\underline{i}}\wedge du\, ,
\nonumber \\
& &
\nonumber \\
\hat{\phi} & = & 0\, ,
\label{eq:SSW}
\end{eqnarray}
and the GFS solutions are\footnote{In order to avoid ambiguities we will
always assume that ${\cal A}_{u}-1<0$ so the solution will always have
the same signature as the asymptotic infinity (when the fields vanish).}
\begin{eqnarray}
\hat{ds}^{2} & = &
2e^{2\hat{\phi}}\{dudv +{\cal A}_{\underline{i}}dx^{\underline{i}}du\}
-dx^{\underline{i}}dx^{\underline{i}}\, ,
\nonumber \\
& &
\nonumber \\
\hat{B} & = & -2e^{2\hat{\phi}}\{(1-e^{-2\hat{\phi}})du\wedge
dv+{\cal A}_{\underline{i}}du\wedge dx^{\underline{i}}\}\, ,
\nonumber \\
& &
\nonumber \\
\hat{\phi} & = & -\frac{1}{2}\log (1-{\cal A}_{u})\, .
\label{eq:GFS}
\end{eqnarray}
Here $i=1,\ldots,8$, $u=\frac{1}{\sqrt{2}}(t+x),\,
v=\frac{1}{\sqrt{2}}(t-x)$, and the fields do not depend on
$x=x^{9}$ and on $t=x^{0}$.

To use the machinery developed in the main body of the paper we have to
identify the nine-dimensional fields. For our purposes it is enough to
do it for just the SSW solutions. First of all we need a zehnbein basis
of the form of Eq.~(\ref{eq:basis}). Fields $k$ and $A_{\mu}$ that
appear in it are readily identified:
\begin{equation}
k = (1-{\cal A}_{u})^{\frac{1}{2}}\, ,\hspace{0.5cm}
A_{t} = \frac{k^{2}-1}{k^{2}}\, ,\hspace{0.5cm}
A_{\underline{i}} = -\frac{1}{\sqrt{2}k^{2}}{\cal A}_{\underline{i}}\, .
\end{equation}
A {\it neunbein} basis, necessary to complete the zehnbein basis, is
provided by
\begin{equation} (e_{\mu}{}^{a})= \left(
\begin{array}{cc}
k^{-1} & 0 \\
-k A_{\underline{i}} & \delta_{\underline{i}}{}^{j} \\
\end{array}
\right)\, ,
\hspace{1cm}
(e_{a}{}^{\mu})=
\left(
\begin{array}{cc}
k & 0 \\
k^{2} A_{\underline{i}} & \delta_{i}{}^{\underline{j}} \\
\end{array}
\right)\, ,
\label{eq:SSWbasis}
\end{equation}
and the rest of the nine-dimensional fields are (with curved indices)
\begin{equation}
\begin{array}{rclrcl}
B_{t\underline{i}} & = & -\frac{1}{2\sqrt{2}}\frac{1+k^{2}}{k^{2}}\,
,\hspace{1.5cm}& B_{t} & = & 0\, ,
\\
& & & & &
\\
B_{\underline{i}\underline{j}} & = & 0\, ,&
B_{\underline{i}} & = & -\frac{1}{\sqrt{2}}{\cal A}_{\underline{i}}
=k^{2}A_{\underline{i}}\, ,
\\
& & & & &
\\
\phi & = & 0\, .& & &
\end{array}
\end{equation}

The field strengths of the nine-dimensional vector fields $A,B$ are
given by
\begin{equation}
\begin{array}{rclrcl}
F_{0i}(A) & = & -2k^{-2}\partial_{\underline{i}}k\, ,&
F_{0i}(B) & = & 0\, ,
\\
& & & & &
\\
F_{ij}(A) & = & -4k^{-1}A_{[\underline{i}}\partial_{\underline{j}]}k
+F_{\underline{i}\underline{j}}(A)\, ,&
F_{ij}(B) & = & k^{2}F_{ij}(A)\, .
\end{array}
\end{equation}

If we write the Killing spinor equation $\delta_{\hat{\epsilon}}
\hat{\psi}_{\underline{x}} =0$ in the form
\begin{equation}
(\partial_{\underline{x}}+M)\hat{\epsilon}=0\, ,
\end{equation}
we have
\begin{equation}
M=\frac{1}{8}\{k^{2}\not\!F(A) -\not\!F(B) -4\hat{\gamma}^{x}
\not\!\partial k \}=\frac{1}{2}(\hat{\gamma}^{0} -\hat{\gamma}^{x})
(\hat{\gamma}^{i}\partial_{\underline{i}}k)\, .
\end{equation}
This implies that the Killing spinor is $x$-independent
$\partial_{\underline{x}} \hat{\epsilon}$ if it is constrained by
\begin{equation}
(\hat{\gamma}^{0} -\hat{\gamma}^{x}) \hat{\epsilon} =0\, .
\end{equation}

This is just the constraint found in Ref.~\cite{kn:BKO}, using a
different, (but also $x$-independent) zehnbein basis, though.  As it was
explained in the end of the previous section, the independence of the
Killing spinor on $x$ in a basis of the form of Eqs.~(\ref{eq:basis}),
(\ref{eq:SSWbasis}) follows from its $x$-independence on any other
$x$-independent frame, in particular that of Ref.~\cite{kn:BKO}.

\item A second example is provided by the dual relation between a
special class of fivebrane solutions \cite{kn:CHS} called
multimonopoles in Ref.~\cite{kn:Kh} and the stringy ALE instantons
\cite{kn:Bian} which have the multicenter Gibbons-Hawking metric.  It
was observed in Ref.~\cite{kn:Bian} that these two solutions are related
by T-duality.  The reason why only the multimonopole configurations are
dual to the stringy ALE instantons is simple.  The characteristic
property of those class of fivebranes is the independence on the
direction $x^{4} = \tau$ which is the one used for duality.  Generic
fivebrane \cite{kn:CHS} as well as generic self-dual metrics
\cite{kn:Bian} do not have such an isometry.  The fivebrane solutions,
including the multimonopoles, have unbroken supersymmetries with
constant chiral (in four-dimensional Euclidean space) Killing spinors in
a $\tau$-independent zehnbein basis.  According to the results of the
previous section this would be sufficient to claim that the dual
solutions (the stringy ALE instantons) have unbroken supersymmetries
with the same Killing spinors.

\item Our last example illustrates our results for time-like duality,
although it cannot be said it is a natural born ``winner".  It is easy
to show that the extreme magnetic dilaton black hole, uplifted to ten
dimensions in \cite{kn:KO2}, is invariant under time-like duality.  We
also know that it has unbroken supersymmetries with constant Killing
spinors restricted by the same condition as the fivebrane Killing
spinors of Ref.~\cite{kn:CHS}: the Killing spinors are chiral in the
four-dimensional Euclidean space spanned by the coordinates
$x^{1},\ldots,x^{4}$, that is
\begin{equation}
(1\pm\hat{\gamma}_{1234})\hat{\epsilon}_{\pm}=0\, .
\end{equation}
Since the configuration is invariant, the Killing spinors {\it are}
invariant too.  On the other hand, in Section~\ref{sec-susy} we found
that the Killing spinors change after time-like duality according to
Eq.~(\ref{eq:dualk2}).  There seems to be a contradiction between these
two facts, but, actually, they are consistent with each other because
the above constraint is invariant under multiplication by
$\hat{\gamma}_{0}(\equiv \hat{\gamma}_{x})$ and the Killing spinor is
simply transformed into another Killing spinor.

It would be interesting to apply our results to supersymmetric
configurations which are not time-like invariant since in general
time-like duality seems to change the sign of the energy of the
configurations and interchanges singularities and horizons
\cite{kn:W,kn:BMQ} while supersymmetry (as we have shown) is preserved.

\end{list}


\section{Conclusion}
\label{sec-conclusion}

Bosonic configurations may have Killing vectors and, when embedded in a
supergravity theory, also Killing spinors. We have studied the case
in which both are present and one performs a T-duality transformation
in the direction associated to a Killing vector.

Usually, the existence of a Killing vector means that there exist a
system of coordinates (adapted coordinates) in which the fields (here
the metric (or zehnbeins), the dilaton, and the two-form field) do not
depend on the coordinate associated to the Killing vector.  One of our
main conclusions is that if a bosonic configuration admits a Killing
vector and a Killing spinor and one uses adapted coordinates, even if
the bosonic fields do not depend on the coordinate associated to the
isometry it is not guaranteed that the Killing spinor will not depend on
it as well.  We have exhibited different examples of this situation.
Our second main conclusion is that in this situation, if one performs a
T-duality transformation in the direction associated to the Killing
vector, the dual configuration will not admit Killing spinors.

The main result of our paper is: T-duality does preserve the unbroken
supersymmetries of those configurations whose Killing spinors are
independent of the coordinate associated to the isometry used for
duality and the Killing spinors transform in a very simple way.

It is interesting to compare this situation with the case of S-duality.
S-duality always preserves the unbroken supersymmetries of the
configurations at the classical level \cite{kn:O}.  However S-duality
and T-duality are on equal footing in some contexts \cite{kn:HuT}: when
the effective action of the type II superstring is compactified on a six
torus, the hidden symmetry of the resulting four dimensional theory
($N=8$ supergravity) is $E_{7}$, which contains the $SO(6,6)$ T-duality
group and the $SL(2,R)$ S-duality group.  Obviously, from the four
dimensional point of view, both T- and S-duality must be consistent with
supersymmetry.  However, in the case of T-duality, we are not interested
in four dimensional configurations for which T-duality amounts to a
rotation of vector and scalar fileds but, often, we are interested in
the nontrivial effects induced by T-duality in the ten-dimensional
metric.  From the ten-dimensional point of view (the one we adopt here)
T-duality will not be consistent with supersymmetry in the cases
explained above.

The investigation of $\alpha^{\prime}$ corrections with respect to
T-duality may also lead to the discovery of some new features.  We know
that T-duality gets $\alpha^{\prime}$ corrections and this means that
the hidden symmetries of the conventional supergravity theories (and the
theories themselves) will be modified in a form unknown at present time.
We have some relevant results on T-duality which includes non-Abelian
vector fields and $\alpha^{\prime}$ corrections which explain the fact
that the SSW \cite{kn:BKO} solutions as well as the dual wave solutions
\cite{kn:BEK} have unbroken supersymmetry with account of
$\alpha^{\prime}$ corrections.  These results will be published
elsewhere \cite{kn:BKO2}.

\section*{Acknowledgements}

We are grateful to E. \'Alvarez, G.W.  Gibbons, G.T.  Horowitz and A.
Tseytlin for most fruitful discussions.  One of us (T.O.) is extremely
grateful to the hospitality, friendly environment and financial support
of the Institute for Theoretical Physics of the University of Groningen
and the Physics Department of Stanford University, where part of this
work was done.  The work of E.B. has been made possible by a fellowship
of the Royal Netherlands Academy of Arts and Sciences (KNAW).  The work
of R.K. was supported by the NSF grant PHY-8612280.  The work of E.B.
and R.K. has also been partially supported by a NATO Collaboration
Research Grant.  The work of T.O. was supported by a European Union
Human Capital and Mobility program grant.


\appendix


\section{From $d=10$ to $d=4$. The supersymmetric truncation}
\label{sec-truncation}

Now we want to make connection with the action of $N=4,d=4$
supergravity.  Therefore we have to compactify six space-like
coordinates and we substitute everywhere $\hat{\eta}_{xx}=-1$.  The
compactification from $N=1,d=10$ to $N=4,d=4$ was done by Chamseddine in
Ref.~\cite{kn:C}.  However Chamseddine worked in the Einstein frame and
that makes very difficult to study the effect of T-duality on the
resultant theory.  Our goal here will be to obtain pure $N=4,d=4$
supergravity (or part of it) in string frame, identifying which fields
belong to the matter multiplet and which fields belong to the
supergravity multiplet and how the dimensionally reduced action has to
be truncated in order to get rid of the matter fields.

We perform the dimensional reduction of the theory from
$d=10$ to $d=4$ for a simplified model in which most of the $d=10$
fields are trivial.  This simplified model is enough to discuss the
important features of the dimensional reduction versus duality.

We do it in two steps.  First we reduce from $d=10$ to $D=5$.  We denote
the $10$-dimensional fields by an upper index $10$ and the
$5$-dimensional fields by a hat.  The $10$-dimensional indices are
capital letters $M,N=0,\ldots,9$, the $5$-dimensional indices will carry
a hat $\hat{\mu},\hat{\nu}=0,\ldots,4$, and the compactified dimensions
will be denoted by capital $I$'s and $J$'s, $I,J=5,\ldots,9$.  We take
the $d=10$ fields to be related to the $D=5$ ones by
\begin{equation}
\begin{array}{rclrcl}
g^{(10)}_{\hat{\mu}\hat{\nu}} & = & \hat{g}_{\hat{\mu}\hat{\nu}}\, ,&
B^{(10)}_{\hat{\mu}\hat{\nu}} & = & \hat{B}_{\hat{\mu}\hat{\nu}}\, ,
\\
& & & & &
\\
g^{(10)}_{I\hat{\nu}} & = & 0\, ,&
B^{(10)}_{I\hat{\nu}} & = & 0\, ,
\\
& & & & &
\\
g^{(10)}_{IJ} & = & \eta_{IJ}=-\delta_{IJ}\, ,\hspace{1.5cm}&
B^{10}_{IJ} & = & 0\, ,
\\
& & & & &
\\
\phi^{(10)} & = & \hat{\phi}\, .& & &
\end{array}
\end{equation}
We get
\begin{equation}
S=\frac{1}{2}\int d^{5}x \sqrt{-\hat{g}} e^{-2\hat{\phi}} [-\hat{R}
+4(\partial\hat{\phi})^{2} -\frac{3}{4}\hat{H}^{2}]\, .
\end{equation}
As a
second step we reduce from $D=5$ to $d=D-1=4$ using the results and
notation of the previous section.  We get
\begin{eqnarray}
S & = &
\frac{1}{2}\int d^{4}x e^{-2\phi}\sqrt{-g} [-R
+4(\partial\phi)^{2} -\frac{3}{4}H^{2}-
\nonumber \\
& &
\nonumber \\
&  & -(\partial\log k)^{2} +\frac{1}{4}k^{2}F^{2}(A)
+\frac{1}{4}k^{-2}F^{2}(B)]\, .
\end{eqnarray}
Now, if we look to the gravitino supersymmetry rule in $d=10$,
\begin{equation}
\delta_{\epsilon}\hat{\psi}_{\hat{c}}=
[\partial_{\hat{c}}-\frac{1}{4}(\hat{\omega}_{\hat{c}}{}^{\hat{a}\hat{b}}-
\frac{3}{2}\hat{H}_{\hat{c}}{}^{\hat{a}\hat{b}})
\hat{\gamma}_{\hat{a}\hat{b}}]\epsilon\, ,
\end{equation}
and observe that setting $k=1$
\begin{equation}
\hat{\omega}_{c}{}^{a4}-\frac{3}{2}\hat{H}_{c}{}^{a4}=
-\frac{1}{2}F_{c}{}^{a}(A+B)\, ,
\end{equation}
it is clear that the identification of the
matter vector fields $D_{\mu}$ and the supergravity vector fields
$V_{\mu}$ is the same as in Chamseddine's paper up to factors of $1/2$:
\begin{eqnarray}
D_{\mu} & = & \frac{1}{2}(A_{\mu}-B_{\mu})\, ,
\nonumber \\
& &
\nonumber \\
V_{\mu} & = & \frac{1}{2}(A_{\mu}+B_{\mu})\, ,
\end{eqnarray}
respectively.
We also have to put $k=1$, because there is no such a scalar in the
$N=4$, $d=4$ supergravity multiplet. Now we want to truncate the theory
keeping only the supergravity vector field $V_{\mu}$. We have then
\begin{equation}
k  =  1\, ,
\hspace{1cm}
V_{\mu}=A_{\mu}=B_{\mu}\, ,
\hspace{1cm}
D_{\mu}=0\, .
\label{eq:truncation}
\end{equation}
The truncated action is
\begin{equation}
S=\frac{1}{2}\int d^{4}x
\sqrt{-g} e^{-2\phi} [-R +4(\partial\phi)^{2} -\frac{3}{4}H^{2}
+\frac{1}{2}F^{2}(V)]\, ,
\label{eq:action4trunc}
\end{equation}
where
\begin{eqnarray}
F_{\mu\nu}(V) & = & 2\partial_{[\mu}V_{\nu]}\, ,
\nonumber \\
& &
\nonumber \\
H_{\mu\nu\rho} & = & \partial_{[\mu}B_{\nu\rho]}+
V_{[\mu}F_{\nu\rho]}(V)\, .
\end{eqnarray}

The embedding of the four-dimensional fields in this action in $d=10$ is
the following:
\begin{equation}
\begin{array}{rclrcl}
g^{(10)}_{\mu\nu} & = & g_{\mu\nu}-V_{\mu}V_{\nu}\, ,\hspace{1.5cm}&
B^{(10)}_{\mu\nu} & = & B_{\mu\nu}\, ,
\\
& & & & &
\\
g^{(10)}_{\underline{x}\nu} & = & -V_{\nu}\, ,&
B^{(10)}_{\underline{x}\nu} & = & V_{\nu}\, ,
\\
& & & & &
\\
g^{(10)}_{\underline{x}\underline{x}} & = & -1\, ,&
\phi^{(10)} & = & \phi\, ,
\\
& & & & &
\\
g^{(10)}_{IJ} & = & \eta_{IJ}=-\delta_{IJ}\, .& & &
\end{array}
\label{eq:uplift}
\end{equation}
This formulae can be used to uplift any four-dimensional field
configuration with one graviton, one axion, one vector and a dilaton to
a ten-dimensional field configuration in a way consistent with
supersymmetry, as in Refs.~\cite{kn:BKO3,kn:KO2}.

One obvious but important observation is that this action is not just
invariant under $x$-duality (here $x=x^{4}$), but {\it all} the fields
that appear in it are individually invariant\footnote{Note that the
truncation itself is duality invariant, i.e. $\tilde{k}=k=1$,
$\tilde{D}_{\mu}=D_{\mu}=0$.}.

But there is more. If we rewrite the truncation
Eq.~(\ref{eq:truncation}) in terms of the original ten-dimensional
fields, it looks like this:
\begin{eqnarray}
g^{(10)}_{\underline{x}\underline{x}} & = & -1\, ,
\nonumber \\
g^{(10)}_{\underline{x}\mu} & = & -B^{(10)}_{\underline{x}\mu}\, ,
\nonumber \\
g^{(10)}_{\underline{x}I} & = & B^{(10)}_{\underline{x}I}=0\, .
\label{eq:truncation4}
\end{eqnarray}

Now one can check that a ten-dimensional configuration satisfying
Eq.~(\ref{eq:truncation4}) is invariant under T-duality in the direction
$x$.  That is also obviously true for the rest of the compactified
directions.

We can state this result as follows: if in the four dimensional action
we interpret the vector field $V_{\mu}$ as belonging to the supergravity
multiplet\footnote{Which is necessary to have supersymmetry.} coming
from the combination $A_{\mu}+B_{\mu}$, then the lifting to ten
dimensions of any four-dimensional configuration will be an $x$-duality
invariant configuration if $x$ is one of the compact dimensions.

One example is provided by the SSW \cite{kn:BKO} and the GFS
\cite{kn:DGHR,kn:BEK} solutions.  These solutions are described in
Section~\ref{sec-examples}, Eqs.~(\ref{eq:SSW},\ref{eq:GFS}).  If
$x=x^{9}$ is the nontrivial compactified dimension (what we called
before $x^{4}$), then, imposing the conditions
Eq.~(\ref{eq:truncation4}) means for these solutions ${\cal
A}_{u}=\hat{\phi}=0$.  This subset of SSW and GFS are identical, are
duality invariant in the $x=x^{9}$ direction and give rise to the same
supersymmetric solutions of $N=4,d=4$ supergravity
\cite{kn:BKO3,kn:KO2,kn:GMGHS,kn:KLOPP}.


\end{document}